\documentclass[a4paper]{article}
\title{Noncommutative quantum mechanics as a constrained system.}
\author{A.A. Deriglazov\footnote{alexei@fisica.ufjf.br ~ On leave of
absence from Dept. Math. Phys., Tomsk Polytechnical University,
Tomsk, Russia.}}
\date{Dept. de F\'\i sica, ICE, Universidade Federal de Juiz de Fora,\\
MG, Brasil.}
\begin{document}
\maketitle
\large
\begin{abstract}
It is shown that quantum mechanics on noncommutative spaces (NQM)
can be obtained by the canonical quantization of some underlying
second class constrained system formulated in extended configuration
space. It leads, in particular, to an intriguing possibility of
quantization in terms of the initial (noncommutative) variables.
Two different formulations are discissed. The first one is appropriate
for at most quadratic potential. The noncommutativity parameter and
rank of matrix of the constraint brackets depend on the potential.
It explains appearance of two phases of the resulting NQM. The second
formulation is appropriate for an arbitrary potential.
In both cases the corresponding
Lagrangian action is presented and quantized, which leads to
quantum mechanics with ordinary product replaced by the Moyal product.
\end{abstract}


\noindent
Recently quantum mechanics on noncommutative spaces (NQM) have received a
considerable interests [1-7]. It can be characterized by the brackets
($\epsilon_{ab}=-\epsilon_{ba}, ~ a, b=1,2, ~ \epsilon_{12}=1$)
\begin{eqnarray}\label{1}
\{x_a, x_b\}=\theta\epsilon_{ab}, \qquad
\{x_a, p_b\}=\delta_{ab}, \qquad
\{p_a, p_b\}=0,
\end{eqnarray}
and by the Hamiltonian
\begin{eqnarray}\label{2}
H=\frac{p^2}{2m^2}+V(x)
\end{eqnarray}
with some potential $V(x)$. To make this situation tractable, the
prescription is to consider the new variables
\begin{eqnarray}\label{3}
\tilde x_a=x_a+\frac{\theta}{2}\epsilon_{ab}p_b, \qquad \tilde p_a=p_a,
\end{eqnarray}
which obey the canonical brackets and thus can be quantized in the
standard way. It leads to the
Schr$\stackrel{..}{o}$dinger equation
\begin{eqnarray}\label{4}
E\Psi(\tilde x)=\frac{1}{2m^2}\tilde p^2\Psi(\tilde x)+
V(\tilde x_a-\frac{\theta}{2}\epsilon_{ab}\tilde p_b)\Psi(\tilde x),
\end{eqnarray}
where the last term can be rewritten [8, 9, 3] through the Moyal product
\begin{eqnarray}\label{5}
V(\tilde x_a-\frac{\theta}{2}\epsilon_{ab}\tilde p_b)\Psi(\tilde x)=
V(\tilde x)\star\psi(\tilde x).
\end{eqnarray}
Thus one obtains quantum mechanics in terms of the commutative variables
$\tilde x, \tilde p$, but  with the ordinary product replaced by the star
product.

Let us recall that in some cases appearance of the noncommutative
geometry [10] has a natural interpretation as
resulting from the canonical
quantization of some underlying constrained system. In particular, this
interpretation is possible for the lowest level Landau problem [11, 9]
and for the open string in a B-field background [12-14]. Concerning the
NQM, one special case was considered in [1, 2], starting from higher
derivative mechanical action. It leads to the noncommutative particle
with extra physical degrees of freedom. In this relation it is naturally
to ask whether a similar interpretation is possible for NQM
(\ref{1})-(\ref{5}). Here we demonstrate that it is actually the case.
Our starting point will be some (ordinary) mechanical system formulated in
an appropriately extended configuration space. Nonphysical character of
the corresponding extra degrees of freedom is supplied by second
class constraints (2CC).
The noncommutative geometry (\ref{1}) for the physical sector variables
arises after introducing of the Dirac bracket, while the
prescription (\ref{3}) becomes, in fact, the standard necessary step
for the canonical quantization of a system with second class constraints
[15, 16].

The results thus obtained are as follows. The Lagrangian action, which is
appropriate for at most quadratic potential, looks as follows
\begin{eqnarray}\label{6}
S=\int d\tau\{-\frac{m}{2}\dot v^2+2(\dot v-\dot x)\theta^{-1}v
-\frac{2}{m\bar\theta^2}v^2-U(x)\},
\end{eqnarray}
where $x_a(\tau), ~ v_a(\tau)$ are the configuration space variables and
$\theta_{ab}=\bar\theta\epsilon_{ab}$. The variables $v_a$ are subject
to 2CC and can be omitted from consideration after the Dirac bracket
introduced. The physical sector consist of $x_a$ and the conjugated
momenta $p_a$. The Dirac bracket for $x_a$ turns out to be nontrivial,
with the noncommutativity parameter being
\begin{eqnarray}\label{8}
\theta=\bar\theta{\det}^{-1}(1-\frac{m\bar\theta^2}{4}\partial\partial U).
\end{eqnarray}
For the case $\partial_a\partial_b U=const$ one can easily find the
canonical variables (see Eq.(\ref{22}) below),
then quantization leads to NQM (\ref{4}) with the potential
$V=-\frac{m\bar\theta^2}{8}\partial_aU\partial_aU+U$. For the case of a
general
potential $U(x)$, the noncommutativity parameter $\theta$ depends on
$x_a$ and one is faced with the problem of diagonalization of the
brackets, Eq.(\ref{20}) below. While existence of the canonical
variables is guaranteed
by the known theorems [16], it is problematic to find a solution in the
manifest form. Surprisingly enough, the
problem can be resolved for an arbitrary potential
if one starts from the action, which is obtained from (\ref{6})
omitting the first term
\begin{eqnarray}\label{9}
S=\int d\tau[2(\dot v-\dot x)\theta^{-1}v-\frac{2}{m\bar\theta^2}v^2-
V(x).
\end{eqnarray}
It can be considered as the action of ordinary particle (with position
$x_a$) written in the first order form, with the
``Chern-Simons'' term for $v$ added: $\dot v\theta^{-1}v$.
The action is similar to the one discussed by Lukierski at all [1], but
do not involves of higher derivatives. As a concequence, there is no
of ``internal'' oscillator modes in the physical sector. Below
we show that this action leads to NQM (\ref{1})-(\ref{5})
with the potential $V=U$.

It is instructive to start from the noncommutative free particle which is
defined by the equations of motion $\dot x_a=\frac{1}{m}p_a, ~ \dot p_a=0$
and by the relations (\ref{1}). To reproduce this system in the framework
of a constrained mechanics, let us consider the following action
\begin{eqnarray}\label{10}
S=\int d\tau\{-\frac{m}{2}\dot v^2+2(\dot v-\dot x)\theta^{-1}v
-\frac{2}{m\bar\theta^2}v^2\}.
\end{eqnarray}
In the Hamiltonian formulation one finds the primary constraints
\begin{eqnarray}\label{11}
G_a\equiv p_a+2\theta^{-1}_{ab}v_b=0,
\end{eqnarray}
and the Hamiltonian
\begin{eqnarray}\label{12}
H=-\frac{1}{2m}\pi^2+\frac{2}{m}\pi\theta^{-1}v+\lambda(p+2\theta^{-1}v).
\end{eqnarray}
Here $p, ~ \pi$ are conjugated momenta for $x, ~ v$ and $\lambda$ is the
Lagrangian multiplier for the constraint. Further analysis leads to
the secondary constraints
\begin{eqnarray}\label{13}
T_{a}=\pi_a-2\theta^{-1}_{ab}v_b=0.
\end{eqnarray}
The constraints form the second class system
\begin{eqnarray}\label{14}
\{G_a, G_b\}=0, \quad \{T_a, T_b\}=-4\theta^{-1}_{ab}, \quad
\{G_a, T_b\}=2\theta^{-1}_{ab},
\end{eqnarray}
and thus can be taken into account by transition to the Dirac bracket
\begin{eqnarray}\label{15}
\{A, B\}_D=\{A, B\}+\{A, G\}\theta\{G, B\}- \cr
\{A, G\}\frac{1}{2}\theta\{T, B\}-
\{A, T\}\frac{1}{2}\theta\{G, B\}.
\end{eqnarray}
Now the variables $v, \pi$ can be omitted from consideration, while
for the remaining physical variables $x, ~ p$ one obtains from
Eq.(\ref{15}) the desired brackets (\ref{1}). To quantize the system
one needs to find the canonical variables [16], which in this case
turn out to be
$\tilde x_a, ~ \tilde p_a$ defined in Eq.(\ref{3}). They obey the standard
brackets
$\{\tilde x_a, \tilde x_b\}=0, ~ \{\tilde x_a, \tilde p_b\}=\delta_{ab},
~ \{\tilde p_a, \tilde p_b\}=0$.
Using Eq.(\ref{11})-(\ref{13}), (\ref{3}) one finds the physical
Hamiltonian $H_{ph}=\frac{1}{2m}\tilde p^2$, which leads
to the free equations
of motion. As it is expected, quantum mechanics of the noncommutative
free particle is identical to the ordinary one.

Let us add some potential\footnote{Short analysis shows
that the choice: $U(y)$ leads to
quantum mechanics with the Hamiltonian
$H_{ph}=\frac{1}{2m}\tilde p^2+V(\tilde p).$}
$U(x)$ to the action (\ref{10}). It leads to deformation of the
constraint algebra, since the secondary constraint involves now
derivative of the potential. Namely, in the Hamiltonian formulation
one has the same primary constraint (\ref{11}), and the Hamiltonian
\begin{eqnarray}\label{16}
H=-\frac{1}{2m}\pi^2+\frac{2}{m}\pi\theta^{-1}v+U(x)+
\lambda(p+\theta^{-1}y).
\end{eqnarray}
Further analysis gives the secondary constraints
\begin{eqnarray}\label{17}
T_{a}\equiv\pi_a-2\theta^{-1}_{ab}v_b+\frac{m}{2}\theta_{ab}\partial_bU=0,
\end{eqnarray}
as well as equations for determining of the Lagrangian multipliers
\begin{eqnarray}\label{18}
F\lambda=-\frac{2}{m}(\pi-\theta^{-1}v),
\end{eqnarray}
where
\begin{eqnarray}\label{19}
F_{ab}\equiv\delta_{ab}-\frac{m\bar\theta^2}{4}\partial_a\partial_b U.
\end{eqnarray}
Next step depends on the rank of the matrix $F$. If $\det F=0$, the model
involves first class constraints (see also Eq.(\ref{20}), which explains
appearance of two phases [4-6] of the resulting NQM. Let us consider the
nondegenerated case $\det F\ne 0$.
Then the constraints form the second class system
\begin{eqnarray}\label{20}
\{G_a, G_b\}=0, \qquad \{T_a, T_b\}=-4\theta^{-1}_{ab}, \qquad
\{G_a, T_b\}=2F_{ac}\theta^{-1}_{cb}.
\end{eqnarray}
Introducing the Dirac bracket
\begin{eqnarray}\label{7}
\{A, B\}_D=\{A, B\}-\triangle^{-1}\{A, G\}\theta\{G, B\}- \cr
\{A, G\}\frac{1}{2}F^{-1}\theta\{T, B\}-
\{A, T\}\frac{1}{2}\theta F^{-1}\{G, B\},
\end{eqnarray}
one obtains the following result for the brackets of the physical variables
\begin{eqnarray}\label{21}
\{x_a, x_b\}=\triangle^{-1}\theta_{ab}, \quad
\{x_a, p_b\}=F^{-1}_{ab},
\quad \{p_a, p_b\}=0.
\end{eqnarray}
The noncommutativity parameter depends now on the potential through the
quantity
\begin{eqnarray}\label{22}
\triangle\equiv\det(1-\frac{m\bar\theta^2}{4}\partial\partial U).
\end{eqnarray}
Let us restrict ourselves to the case $\partial_a\partial_b U=const$.
Then the canonical variables can be defined as
\begin{eqnarray}\label{22}
\tilde x_a=F_{ab}x_b+\frac{1}{2}\theta_{ab}p_b, \quad
\tilde p_a=p_a.
\end{eqnarray}
The Hamiltonian in terms of the canonical variables is
\footnote{The term $F$ can be equally included into the kinetic part
of the Hamiltonian. To achieve this one defines the canonical variables as
\begin{eqnarray}
\tilde x=x+\frac{1}{2\triangle}\theta Fp, \quad
\tilde p=Fp. \nonumber
\end{eqnarray}
Then the  Hamiltonian is
\begin{eqnarray}
H_{ph}=\frac{1}{2m}\tilde pF^{-2}\tilde p-
\frac{m\bar\theta^2}{8}\partial_aU\partial_aU+
U(\tilde x-\frac{1}{2\triangle}\theta\tilde p).\nonumber
\end{eqnarray}
Note that for the Galilean invariant system $F$ is proportional to the
unit matrix.}
\begin{eqnarray}\label{23}
H_{ph}=\frac{1}{2m}\tilde p^2-
\frac{m\bar\theta^2}{8}\partial_aU\partial_aU|_{x(\tilde x, \tilde p)}+
U[F^{-1}(\tilde x-\frac{1}{2}\theta\tilde p)],
\end{eqnarray}
where the term with derivatives of the potential comes from Eq.(\ref{17}).
The resulting system can be quantized now by the standard way. Note that
the underlying potential $U$ and the final one turn out to be different
for this model. For example, starting from the harmonic oscillator
$U=\frac{k}{2}|x|^2$, one obtains the NQM which corresponds to
oscillator with
renormalized rigidity $\tilde k=(1-\frac{m\bar\theta^2k}{4})^{-1}k$,
namely
\begin{eqnarray}\label{24}
V=
[-\frac{m\bar\theta^2}{8}\partial_aU\partial_aU+
U]|_{x(\tilde x, \tilde p)}=\frac{\tilde k}{2}
|\tilde x-\frac{1}{2}\theta\tilde p|^{2}.
\end{eqnarray}

Let us return to the case of an arbitrary potential. As it was mentioned,
the complicated brackets (\ref{21}) arise due to the fact that the
secondary constraints (\ref{17}) involve derivative of the potential.
One possibility to avoid the problem is to construct action which will
create the primary constraints only. Since $U(x)$ do not contains of
the time derivative, it can not give contribution into the primary
constraints. An appropriate action is\footnote{An equivalent form of
the action can be obtained by the shift: $x \to x'=x-v$.}
\begin{eqnarray}\label{25}
S=\int d\tau[2(\dot v-\dot x)\theta^{-1}v-\frac{2}{m\bar\theta^2}v^2-
V(x),
\end{eqnarray}
where $x_a, ~ v_a$ are the configuration space variables. Configuration
space dynamics is governed by second order equations which is supplied
by the term $v^2$.
Following the Dirac procedure one obtains primary second
class constraints
\begin{eqnarray}\label{26}
G_a\equiv p_a+2\theta^{-1}_{ab}v_b=0, \qquad
T_{a}=\pi_a-2\theta^{-1}_{ab}v_b=0,
\end{eqnarray}
and the Hamiltonian
\begin{eqnarray}\label{27}
H=\frac{2}{m\bar\theta^2}v^2+V(x)+\lambda_1G+\lambda_2T.
\end{eqnarray}
The constraints are the same as for the free particle, see Eq.(\ref{11}),
(\ref{13}), so the remaining analysis is similar to that case. After
introducing of the Dirac bracket (\ref{15}) the variables $v, ~ \pi$ can
be omitted, while for $x, ~ p$ one has the brackets (\ref{1}). Defining
the canonical variables
\begin{eqnarray}\label{28}
\tilde x_a=x_a+\frac{1}{2}\theta_{ab}p_b, \qquad \tilde p_a=p_a,
\end{eqnarray}
one obtains the physical Hamiltonian
\begin{eqnarray}\label{29}
H=\frac{2}{m}\tilde p^2+V(\tilde x-\frac{1}{2}\theta\tilde p),
\end{eqnarray}
thus reproducing the NQM (\ref{4}), (\ref{5}) for the case of
arbitrary potential.

In conclusion, we have demonstrated that quantum mechanics on
noncommutative space can be considered as resulting from direct canonical
quantization of some underlying constrained system. It implies, that
instead of the star product (\ref{4}), (\ref{5}), one can equally use now
other possibilities to quantize the system. In particular, the
conversion scheme [17] or the embedding formalism [18] can be applied.
For example, it is not difficult to rewrite the formulation (\ref{25})-
(\ref{27}) as a first class constrained system. Namely, let us keep
$G$-constraint only and define the deformed Hamiltonian as
\begin{eqnarray}\label{30}
\tilde H=\frac{2}{m\bar\theta^2}v^2+V[x-\frac{1}{2}\theta
(\pi-2\theta^{-1}v)]+\lambda G.
\end{eqnarray}
Since $\{G, \tilde H\}=0$, it is equivalent formulation of the problem
(\ref{27}), the latter is reproduced in the gauge $T=0$. Now one can
quantize all the variables canonically, while the first class constraint
$G=0$ can be imposed as restriction on the wave function. It implies
quantization in terms of the initial noncommutative variables.
Other possibility is to consider the gauges different from $T=0$.
For example, one can take $\pi=0$, which can lead to simplification
of the eigenvalue problem (\ref{4}). Let us point also that the action
(\ref{25}) can be easily generalised on three dimensional case. These
problems will be considered elsewhere.

\section{Acknowledgments.}
Author thanks INFN for the financial support and hospitality during
his stay in Frascati, Italy, where part of this work was done.

\end{document}